\documentclass{elsart5p}

\usepackage[usenames]{color}
\usepackage{graphicx}% Include figure files
\usepackage{enumerate}
\usepackage{ifpdf}
\usepackage[colorlinks]{hyperref}
\usepackage{amsmath}
\usepackage{amssymb}
\usepackage{epsfig}
\usepackage{natbib}
\def\pbar{$\overline{p}$}
\def\dbar{$\overline{d}$}
\begin{document}
%\preprint{LAPTH}
%\input epsf

\begin{frontmatter}
\title{Transport of exotic anti-nuclei: II- $\overline{p}$ and $\overline{d}$ astrophysical uncertainties}

\author[INFN,LPNHE]{D. Maurin\corauthref{cor}}
\ead{dmaurin@lpnhe.in2p3.fr}
\author[LAPTH,univ]{R. Taillet}
%\ead{taillet@lapp.in2p3.fr}
\author[DIAS,LAOG]{C. Combet},
%\ead{ccombet@obs.ujf-grenoble.fr}

\address[INFN]{Dipartimento di Fisica Teorica, Universit\`a di Torino, Istituto Nazionale
di Fisica Nucleare, via P. Giuria 1, I--10125 Torino, Italy}
\address[LPNHE]{LPNHE, IN2P3/CNRS/Universit\'es Paris VI et Paris VII, 4 place Jussieu
Tour 33 - Rez de chauss\'ee, 75252 Paris Cedex 05, France}
\address[LAPTH]{Laboratoire de Physique Th\'eorique {\sc lapth},
             Annecy--le--Vieux, 74941, France}
\address[univ]{Universit\'e de Savoie, Chamb\'ery, 73011, France}
\address[DIAS]{Dublin Institute for Advanced Studies {\sc dias},
       5 Merrion Square, Dublin 2, Ireland}
\address[LAOG]{Laboratoire d'Astrophysique, Observatoire de Grenoble {\sc laog},
			BP 53 F-38041 Grenoble Cedex 9, France}
\corauth[cor]{Corresponding author.}

\begin{abstract}
We use a 1D propagation model to study the dependence of the \pbar\ and \dbar\ exotic
fluxes on the transport parameters. The simple analytical solutions allow us
i) to clarify the origin of the astrophysical uncertainties, and ii) to compare
two models used for {\em signal} predictions, namely the constant and the linear
Galactic wind models. We also study how these uncertainties should be reduced
using forthcoming nuclear cosmic ray data. We confirm that the degeneracy of the transport
parameters for a given propagation model leads to very different fluxes for primary antinuclei
$(\sim 10^2)$. However, we show that with forthcoming data, these uncertainties could be
greatly reduced $(\sim 2)$. As the precision will increase, the astrophysical uncertainty
could then be dominated by our ignorance of the correct spatial dependence for some of the
transport parameters: for instance, the constant and the linear wind models do
not predict the same amount of exotic \pbar\ at low energy.
\end{abstract}


\end{frontmatter}

%\pacs{98.38.Cp, 98.35.Pr, 96.40.-z, 98.70.Sa, 96.50S-, 96.50sb, 95.30.Cq, 12.60.Jv, 95.35.+d}
% 98.38.Cp Milky Way 
% 98.35.Pr Solar neighborhood
%---
% 96.40.-z Cosmic rays (in the solar system)
% 96.50.S- Cosmic rays
% 96.50.sb energy spectra, 
% 98.70.Sa Cosmic rays (outside the solar system)
%---
% 95.30.Cq Elementary particle processes in astrophysics
% 12.60.Jv Supersymmetric models, 
%---
%Dark matter, 95.35.+d

\maketitle
%\tableofcontents

%/////////////////////////////////////////////////////////////////
\section{Introduction}
\label{sec:intro}
Positron, anti-proton and anti-deuteron fluxes are among the primary targets
of ongoing ({\sc bess, pamela}) and forthcoming experiments ({\sc ams}, {\sc gaps}),
seeking for indirect hints at dark matter candidates. 
These particles, unlike $\gamma$-rays, are very sensitive to the
random magnetic fields pervading the Galaxy, and their
transport is diffusive in nature.

Several propagation models
(e.g., {\sc DarkSusy}---\citealt{2004JCAP...07..008G};
{\sc Galprop}---\citealt{2005AdSpR..35..156M};
or \citealt{2001ApJ...563..172D})
are being used to calculate the standard background fluxes
as well as possible exotic contributions from new particles
filling the dark matter halo. 
These calculations aim to
uncover excesses over the background, or to
put some constraints on parameters of new physics
if none is observed. 
In this context, it is crucial to understand the uncertainties
associated with the background and the hypothetical signal.
Throughout this paper, we focus on astrophysical uncertainties
only. To estimate them, we assume that the source
spectrum and spatial distribution of the exotic
matter are fixed, leaving the propagation coefficients as the
only free parameters of flux calculations.

Whatever the propagation
model, the exotic fluxes (and astrophysical uncertainties)
depend on the associated transport coefficients (and their uncertainties).
The usual way to derive all these quantities is as follows:
\begin{enumerate}[1.]
 \item Transport parameters are determined using secondary to primary ratios,
usually B/C, i.e. from  {\em standard} GCRs
whose sources are located {\em in the disk} of the Galaxy.
A numerical study is generally required to extract
         \begin{enumerate}
             \item the parameters yielding the best fit to data;
             \item uncertainties on these parameters.
          \end{enumerate}
\item These parameters are then used to calculate
   \begin{enumerate}
       \item $e^+$, \pbar, and \dbar\ secondary fluxes (background),
       whose sources are also in the Galactic disk;
       \item $e^+$, \pbar, and \dbar\ exotic fluxes\footnote{Below, we use
       indifferently 'exotic' or 'primary' fluxes.},
       whose sources are now located in {\em the diffusive halo}
       of the Galaxy.
   \end{enumerate}
Transport coefficients from the best fit give the most likely
$e^+$ (or \pbar\, $\bar{d}$)
fluxes, whereas the range of allowed transport parameters determine
their uncertainties (for the chosen modelling).
\end{enumerate}

A thorough study along the lines sketched above has been performed
in the framework of the 2D constant wind diffusion model
\citep{2001ApJ...555..585M,2001ApJ...563..172D,2005PhRvD..72f3507B}.
With respect to item (1), \citet{2001ApJ...555..585M} found that
the set of allowed propagation parameters is strongly degenerated.
As regards (2a), \citet{2001ApJ...563..172D} found that this degeneracy
has almost no
consequence on the secondary \pbar\ (or $\bar{d}$)  fluxes (uncertainties are of
about 25\% at most)---which is expected for any modelling, as long
as the propagation parameters are adjusted from B/C data.
Finally, related to (2b), \citet{2005PhRvD..72f3507B}
showed that this degeneracy leads
to an uncertainty on exotic \pbar\ fluxes of about two orders
of magnitude at low energy.
The purpose of this paper is partly to explicit with simple
calculations the above result, then to extend the analysis
to different propagation models.

The outline of the paper is the following: in Sec.~\ref{sec:preamble}, we start
with a description of the 1D modelling used throughout the paper.
Section~\ref{sec:degener} deals with astrophysical uncertainties
of \pbar\ and \dbar\ primary exotic fluxes. The analysis is performed
for the case of pure diffusion (\S\ref{sec:diff}), then for the case
of diffusion/convection models (\S\ref{sec:1Dlinear-wind}). The predictions
for the constant and the linear wind models are compared in Sec.~\ref{sec:generalization},
using the best-fit values of the transport parameters provided in the
literature. Finally, in Sec.~\ref{sec:red_uncer}, we focus on the
constant wind model and estimate by how much the existing uncertainties
on the \pbar\ and \dbar\ exotic fluxes could be reduced with better
data.

%/////////////////////////////////////////////////////////////////
\section{1D diffusion models}
\label{sec:preamble}
Diffusion models in which the density is assumed to depend only
on the $z$-coordinate (1D models) provide a simple but useful
description of the propagation of cosmic ray nuclei
\citep{1979ApJ...229..747J,2001ApJ...547..264J}.

Regarding exotic sources, we showed in the companion paper \citep{Maurin:2006hy}
that analytical 1D--models provide a fair description of the \pbar\ primary fluxes, compared
to that calculated from a full 2D--model taking into account energy gain and losses.
The main reason is that the energy redistribution terms are restricted to the thin
disk. However, exotic sources of \pbar\ are located in the extended diffusive halo
of the Galaxy, and the \pbar\ reaching us rarely cross the disk on average.

Thus, if we discard energy gains and losses and use a 1D geometry, the \pbar\
and \dbar\ flux calculations will still be satisfactory for a gross estimate
(compared with full 2D calculation), but now the advantage is that we
deal with simple analytical solutions (see App.~\ref{App:1D-standard}
and~\ref{App:1D-exotic}). These solutions allow us to put to the fore the
key parameters driving exotic anti-nuclei uncertainties, as well as
a their quick estimates.

\subsection{Short description}
  \label{App:general-case}
\begin{figure}[!t]
\begin{center}
\includegraphics[width=\columnwidth]{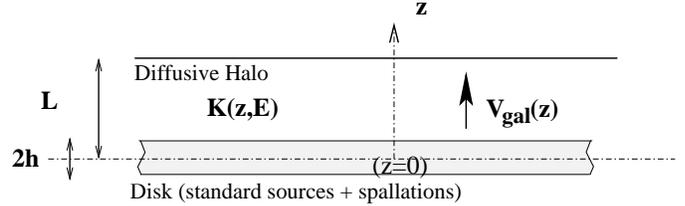}
\caption{\small Upper half-plane of the 1D infinite disk along $r$
(half-thickness $h$). Cosmic rays diffuse in the disk and in the halo,
with a diffusion coefficient $K(z,E)$.}
\label{fig:Lbound2}
\end{center}
\end{figure}

We consider an infinite plane whose density and source distribution
do not depend on $r$ (see Fig.~\ref{fig:Lbound2}).
The thin-disk approximation is used ($h\ll L$).
The diffusion/convection transport equation (without energy redistributions)
for a constant source term $q_{\rm Dark}(E)$ reads
\begin{eqnarray}
\label{1D-generic}
 -\!\frac{d}{dz}\!\left\{\!K(z)\frac{dN}{dz}\!\right\}\!
 \!+\!\frac{d}{dz} [V_{\rm gal}(z) N]
 \!+\! 2h \Gamma \delta(z) N \!=\!q_{\rm Dark}(E).
\end{eqnarray} 
The spatial and energy dependence of the diffusion coefficient
are assumed to be independent: $K(z,E)=\beta K_0(z) {\cal R}^{\delta}$
($\cal R$ is the rigidity, $\beta=v/c$).

Discussions in the paper will refer to the four propagation parameters of the model:
i) the halo size of the Galaxy $L$, ii) the diffusion coefficient normalization
$K_0(z)$, iii) the diffusion slope $\delta$, and iv) the value of the Galactic
wind $V_{\rm gal}(z)$.
Many propagation models are
equivalent to a Leaky Box Model (LBM), as long as stable
species originating from the disk are considered~\citep{1970PhRvD...2.2787J}.
Hence, transport parameters of 1D--models can be easily inferred
(step 1 described in Sec.~\ref{sec:intro})
from the mean grammage $\langle x\rangle^{LB}=\lambda_{\rm esc}(E)$
of the LBM:
the details of this equivalence as well
as specific relationships are reminded in App.~\ref{app:LBvsDM};
mean grammage in various 1D configurations are given in
App.~\ref{App:1D-standard}.
Step (2) of the analysis, i.e. estimating uncertainties on exotic fluxes,
is then straightforward using 1D formulae derived in App.~\ref{App:1D-exotic}.

For the sake of simplicity, spallative destruction of exotic species in the disk is
neglected throughout the paper. It is checked in App.~\ref{app:nospal}, for the constant
wind model, that this leaves the conclusions unchanged.
This is understood as the bulk of exotic \pbar\ reaching us are mostly those created
in the halo: on average, their disk crossings are too rare to yield a sizeable
effect on the low energy spectrum (energy redistribution are neglected for the same reason).

\subsection{Type I and Type II uncertainties}
Two types of uncertainties are identified and discussed in this paper:
 \begin{description}
               \item {\em Type I (parameter uncertainty)} corresponds to uncertainties related to 
							 the degeneracy
               of transport parameters for a given propagation model (i.e. departure from best fit).
               \item {\em Type II (modelling uncertainty)} corresponds to the different predictions of primary fluxes
               between different propagation models whose
               respective transport parameters give the same standard fluxes.
       \end{description}
We focus on the following configurations: pure diffusion (with one diffusion coefficient)
and constant or linear wind model.

\section{Degeneracy in the transport parameters}
\label{sec:degener}
The diffusion coefficient is assumed independent of the spatial coordinates,
so that $K_0(z)\equiv K_0$.
                   %---------------%

       \subsection{Pure diffusion}
       \label{sec:diff}

                        %%%%
       \subsubsection{Degeneracy of parameters for standard sources}
In the thin disk model, the transport equation reads:
\[
 -KN''+nv\sigma 2h\delta(z) \times N = 2h\delta(z) q.
\]
For $z=0$, the solution is
       \[
               N(0)=\frac{2hq}{ 2K/L + nv\sigma}\;.
       \]
Inserted in Eq.~(\ref{<x>2}), this gives for the mean grammage
\begin{equation}
               \label{pure-DM}
               \langle x\rangle^{pure-DM} =\frac{\mu vL}{2K}\;,
       \end{equation}
where $\mu\equiv 2hn\bar{m}$ is the surface mass density in the disk, $L$ the halo
size, $K$ the diffusion coefficient and $v$ the velocity of the nucleus.

As the mean grammage in the LB model is given by
$\langle x\rangle^{LB}=\lambda_{\rm esc}(E)$,
the diffusion coefficient is inferred from Eq.~(\ref{x-equiv})--- i.e.
$\langle x\rangle^{pure-DM}=\langle x\rangle^{LB}$---and
Eq.~(\ref{pure-DM}):
       \begin{equation}
			 \label{eq:titi}
               K(E) = \frac{\mu vL}{2\lambda_{\rm esc}(E)}\;.
       \end{equation}
Both the diffusion model and the leaky box model
lead to the same B/C ratio as long as Eq.~(\ref{eq:titi}) holds.
Consequently, both models lead to the same standard secondary fluxes
for \pbar\ and \dbar.

                               %%%%
               \subsubsection{Lifting the degeneracy: uncertainties for primary \pbar\ and \dbar\
               in the halo}
               \label{uncert-TypeI}
In contrast, exotic sources are
located in the whole diffusive halo, which lifts the above mentioned
degeneracy. The equation (no spallations) now reads
\[
 -KN'' = q_{\rm Dark},
\]
whose solution is given by Eq.~(\ref{prim:pureDiff-nospal})
\begin{equation}
       \label{prim:pure-Diff}
 N_{(\bar{p},\;\bar{d})}^{pure-DM}(z=0)= \frac{q_{Dark}L^2}{2K}=\frac{q_{Dark}\lambda_{\rm esc}L}{\mu
 v}\;.
\end{equation}

For a given value of $\lambda_{\rm esc}$, i.e. a given B/C ratio, the
primary flux depends linearly\footnote{This behaviour can be seen in Fig.~8 of \citet{2002A&A...388..676B} for 
a 2D model using a realistic dark matter distribution (for PBH sources).} on $L$. 
If a very conservative range, $1-15$~kpc, is assumed for $L$ (as taken for
example in~\citealt{2001ApJ...555..585M}), then
the primary flux is uncertain by a factor of 15.

                         %---------------%
       \subsection{Adding a constant wind $V_{\rm gal}=V_c$}
       \label{sec:1Dlinear-wind}
Keeping the same spatial dependence for the diffusion coefficient,
a constant wind directed outward the Galaxy
is now added (pure diffusion recovered when $V_c\to 0$).

One of the main appeal for adding a constant wind comes from the
measured form of $\lambda_{\rm esc}(E)$. A fit to the B/C data in the Leaky
Box Model---including all ingredients, i.e. energy losses, etc.---
gives (see, e.g., \citealt{1998ApJ...508..940W})
  \begin{eqnarray}
  \label{Forme lambda}
        \lambda_{\rm esc}({\cal R})=
   \left\{
    \begin{array}{l}
      \beta\times 15.61 {\rm ~g~cm}^{-2} \mbox{~~~~if~${\cal R}<3.6\,\text{GV}\,$;}\vspace{0.15cm}\\
      \beta \left( {\cal R}/3.6 \right)^{-0.7}\!\!\times\!15.61 {\rm ~g~cm}^{-2}  \mbox{~otherwise.}
    \end{array}
    \right.
  \end{eqnarray}
Such a form naturally arises
in diffusion/convection models (compare $\langle x \rangle^{V_c}$ and $\lambda_{\rm esc}(E)$
at low energy). This remarkable result was first pointed
out by \citet{1979ApJ...229..747J}: the phenomenological quantity
$\lambda_{\rm esc}(E)$, taken from the LBM, hints at a dual transport in
propagation models (see also \citealt{2001ApJ...547..264J} for other
interpretations). Beyond a few GV, the shape of $\lambda_{\rm esc}$ shows that
diffusion always prevails so that conclusions of Sec.~\ref{sec:diff} still hold
at high energy.

The mean grammage is obtained from Eq.~(\ref{x-1D-Vc-thin})
       \begin{equation}
			 \label{eq:re-grammage_Vc}
               \left<x\right>^{V_c}\equiv
               \frac{\mu v}{2V_c} \left[1-e^{-\frac{V_cL}{K}}\right]\;.
       \end{equation}
This formula has two asymptotic behaviours with energy
($K(E)$ depends on energy): i) a diffusion-dominated regime if
$V_cL/K\ll 1$ (reached at high energy) and ii) a convection-dominated
regime if $V_cL/K\gg 1$ (low energy).
The limiting cases are
\begin{equation}
       \label{eq-lim:x-Vc-DM}
 \langle x \rangle^{V_c} \; \stackrel{V_c \ll  \frac{K}{L}}{\rightarrow}
 \; \frac{\mu vL}{2K}
 \quad \text{and} \quad
 \langle x \rangle^{V_c} \; \stackrel{V_c \gg \frac{K}{L}}{\rightarrow}
 \; \frac{\mu v}{2V_c}\,.
\end{equation}
At high energy, the same interplay between the propagation
parameters applies as in Sec.~\ref{sec:diff}.
At low energy, an upper limit for the wind velocity
is found when the convection-dominated regime is reached:
equating rhs of Eq.~(\ref{eq-lim:x-Vc-DM}) and first line of Eq.~(\ref{Forme lambda}),
with a standard value $\mu=2.4\times10^{-3}$~g~cm$^{-2}$ gives,
       \begin{equation}
               \label{Vc}
               V_c \sim 20 {\rm ~km~s}^{-1}.
       \end{equation}

Equation~(\ref{prim:pureDiff-nospal}) gives the flux
for exotic sources in the galactic halo (as before, spallations neglected):
       \begin{equation}
               \label{prim:Vc-Diff}
               N_{(\bar{p},\;\bar{d})}^{V_c}(z=0)= \frac{q_{\rm Dark}K}{V_c^2}
               \left[ 1-e^{-\frac{V_cL}{K}}
               \left(1\!+\!\frac{V_cL}{K}\right)\right]
       \end{equation}
which asymptotic forms read
\begin{equation}
       \label{eq-lim:prim-Vc-DM}
 ~\!\!\!\!N_{(\bar{p},\;\bar{d})}^{V_c} \stackrel{V_c \ll  \frac{K}{L}}{\rightarrow}
 \; \frac{q_{\rm Dark}L^2}{2K}
 \!\quad\!\! \text{and} \!\!\quad\!
 N_{(\bar{p},\;\bar{d})}^{V_c} \stackrel{V_c \gg  \frac{K}{L}}{\rightarrow}
 \; \frac{q_{\rm Dark}K}{V_c^2}.
\end{equation}

Equation~(\ref{eq:re-grammage_Vc}) gives $V_cL/K$ as a function of $\lambda_{\rm esc}$,
provided that $V_c < \mu v/2\lambda_\text{esc}$ ($\sim$ 20~km~s$^{-1}$).
Equation~(\ref{prim:Vc-Diff}) is rewritten
\begin{equation}
  N_{(\bar{p},\;\bar{d})}^{V_c}(0)= \frac{4q_{\rm Dark} \lambda_\text{esc}^2 K}{\mu^2 v^2}
  \frac{\left[ 1-e^{-\frac{V_cL}{K}}\left(1\!+\!\frac{V_cL}{K}\right)\right]}{\left[1-e^{-\frac{V_cL}{K}}\right]^2}\;.
\end{equation}
For a fixed value of $\lambda_\text{esc}$, the exotic flux depends
linearly on $K$ and only weakly on the combination $V_c L/K$.
This gives an overall uncertainty similar to the asymptotic case.

                       \subsection{Linear wind $V_{\rm gal}(z)=V_l\times z$}
               \label{subsub:lin-cte}
A linear wind $V_{\rm gal}(z)=V_l\times z$ is now considered ($[V_l]=$km~s$^{-1}$~kpc$^{-1}$).
The mean grammage is obtained from Eq.~(\ref{x-1D-Vl-thin})
\begin{equation}
 \langle x \rangle^{V_l} \equiv\frac{\mu v \sqrt{\pi}\; {\rm erf} \left( \sqrt{\frac{V_l}{2K}}L\right)}{
       4K\sqrt{\frac{V_l}{2K}}}
\end{equation}
and the primary flux is given by Eq.~(\ref{prim:Vl-nospal})
\begin{equation}
       \label{prim:Vl-Diff}
 N_{(\bar{p},\;\bar{d})}^{V_l}(0)= \frac{q}{V_l}\left(1-e^{-\frac{V_lL^2}{2K}}\right)\;\;.
\end{equation}
The corresponding limiting cases are
\begin{equation}
       \label{eq-lim:x-Vl-DM}
 \langle x \rangle^{V_l} \; \stackrel{V_l \ll  \frac{K}{L^2}}{\rightarrow}
       \; \frac{\mu vL}{2K}
 \quad \text{and} \quad
 \langle x \rangle^{V_l} \; \stackrel{V_l \gg \frac{K}{L^2}}{\rightarrow}
       \; \frac{\mu v\sqrt{\pi}}{\sqrt{8KV_l}}\;,
\end{equation}
and
\begin{equation}
       \label{eq-lim:prim-Vl-DM}
 N_{(\bar{p},\;\bar{d})}^{V_l} \; \stackrel{V_l \ll  \frac{K}{L^2}}{\rightarrow}
 \; \frac{qL^2}{2K}
 \quad \text{and} \quad
 N_{(\bar{p},\;\bar{d})}^{V_l} \; \stackrel{V_l \gg  \frac{K}{L^2}}{\rightarrow}
 \; \frac{q}{V_l}\;.
\end{equation}

At low energy, we proceed as in the previous section
(at high energy, the pure diffusive
transport is recovered).
It is first assumed that the convective transport regime
$V_l \gg  K/L^2$ holds and that the mean-grammage is perfectly
determined from hypothetical perfect data. The
mean grammage fixes $V_l$ through right-hand side Eq.~(\ref{eq-lim:x-Vl-DM}).
This time, the degeneracy on $K_0/L$ leads to no uncertainty on the exotic flux, 
at variance with the former case:
\[
 N_{(\bar{p},\;\bar{d})}^{V_l} \approx \frac{q}{V_l}
               \quad {\rm compared~to}  \quad
N_{(\bar{p},\;\bar{d})}^{V_c} \approx \frac{qK}{V_c^2} \;.
\]
This simply pertains to the absence of the factor $L$
or $K$ in the left-hand side formula. 

As in Sec.~\ref{sub:naiveII}, we can
derive estimates of the range span by $V_l$ from real data.
At low energy, one should obtain from Eqs.~(\ref{Forme lambda})
and~(\ref{eq-lim:x-Vl-DM}) the relation
\[
15.61 {\rm ~g~cm}^{-2} =  \frac{\mu c\sqrt{\pi}}{\sqrt{8KV_l}}\;.
\]
From the range of variation
of the LB model (i.e. $\delta=0.3-0.7$)
using $\mu=2.4\times 10^{-3}$~g~cm$^{-2}$, we get
\[
V_l \sim 8-30 {\rm ~km~s}^{-1}{\rm ~kpc}^{-1}.
\]
This value is not too sensitive to the uncertainty in $L$ because the latter
translates in $V_l$ through $\sqrt{K}$ in the mean-grammage formula above.
This is compatible with results gathered in Tab.~2 of \citet{2005JCAP...09..010L}
(for $\delta=0.3$). The ``parameter" uncertainties (Type I)
for this model are moderate (a factor of 3 here) and smaller
than at high energy: it was the reverse for the constant wind model.

                         %-----------------%
       \section{Difference between models: constant vs linear wind}
       \label{sec:generalization}

The constant wind and the linear wind models are
both used to exclude (or fit) parameters of new physics
\citep{2004PhRvD..69f3501D,2005JCAP...09..010L}. It is thus of paramount
importance to check if they predict the same amount of exotic \pbar\ and $\bar{d}$.

This difference exists only at low energy when the two models differ from pure diffusion.
Taking the ratio of the
primary flux formula of the linear-wind model Eq.~(\ref{prim:Vl-Diff}) to that of
the constant-wind model Eq.~(\ref{prim:Vc-Diff}) leads to
\[
 \frac{N_{(\bar{p},\;\bar{d})}^{V_l}}{N_{(\bar{p},\;\bar{d})}^{V_c}}
 = \frac{V_c^2}{K_cV_l} \times \frac{\left(1-e^{-\frac{V_lL^2}{2K_l}}\right)}{
               \left[ 1-e^{-\frac{V_cL}{K_c}}\left(1\!+\!\frac{V_cL}{K_c}\right)\right]}\;,
\]
where $K_l$ and $K_c$ are the value of the diffusion coefficients found
respectively in the linear and constant wind model. As a new exotic component
is usually seeked at low energy, we only calculate this quantity for
a kinetic energy $E_k\sim 600$~MeV (IS), corresponding roughly to
${\cal R}\sim 1$~GV, so that $K(1~{\rm GV})\sim K_0$.

The diffusion coefficient $K$ should be constrained to be the same in both
models (high energy constraint). Table~2 of~\citet{2005JCAP...09..010L}
gives $K_0=2.5\times 10^{28}$~cm$^2$~s$^{-1}$, $L=4$~kpc and $V_l=6$~km~s$^{-1}$~kpc$^{-1}$
(for $\delta=0.55$) whereas for the constant wind model, Tab.~II of~\citet{2005PhRvD..72f3507B}
gives $K_0=1.35\times 10^{28}$~cm$^2$~s$^{-1}$, $L=4$~kpc and $V_c=12$~km~s$^{-1}$ (for $\delta=0.7$).
Note that both best fits correspond to $L=4$~kpc (although it could be only a coincidence).
This minimizes the difference that is observed between the two calculations
as different $L$ give different primary fluxes (see Sec.\ref{uncert-TypeI}).
Plugging these parameters in the previous formula gives
\[
 \frac{N_{(\bar{p},\;\bar{d})}^{V_l}}{N_{(\bar{p},\;\bar{d})}^{V_c}}\sim 1.3\;.
\]
This is not striking a difference, but this figure is expected to increase
at lower energy. Moreover, it would also be larger would the best-fit value for $L$ in
the two models be different. 

In any case, we remind that in any models, the transport parameters are more or less
degenerate, providing the right amount of B/C. However, as explained in a previous section,
this degeneracy is broken for (exotic) sources in the Galactic halo. The way it is broken
depends on the vertical dependence of the transport parameters. Hence, different models
are expected to lead to different predictions for the exotic flux. This issue
deserves a dedicated study, which goes beyond the scope of this first simple
discussion.

%/////////////////////////////////////////////////////////////////
\section{Reducing uncertainties}
\label{sec:red_uncer}
                        %%%%
               \subsection{Impact of the inaccuracy of B/C data}
\label{sub:naiveII}
So far, we discussed the hypothetical case of ideal data to estimate
the degeneracy in the transport parameters. Using real ones, a greater
degeneracy is generally obtained, enhancing the uncertainties on 
the exotic fluxes.

In LB models, the acceptable range of parameters found in
the literature providing a good fit to B/C is $\delta=0.3-0.7$.
This value of the spectral index in $\lambda_{\rm esc}({\cal R})$ 
[see Eq.~(\ref{Forme lambda})] depends on the level of
reacceleration \citep{2001ApJ...547..264J}.
This translates into an increase of the range spanned by
$\lambda_{\rm esc}({\cal R})$, and thus to an increase
of the propagation parameter range extracted from it in the 1D model.
\begin{table}
\centering
\begin{tabular}{c c c  c c}
\hline\hline
 ~~~~Set~~~~& $\delta$ & ~$K_0$ (kpc$^2$~Myr${^{-1}}$)~ & $L$ (kpc)& ~$V_c$ (km~s$^{-1}$) \\ \hline
{\em max} & 0.46 & 0.0765 & 15 & 5\\
{\em best} & 0.7 & 0.0112 & 4 & 12 \\
{\em min} & 0.85  & 0.0016 & 1 & 13.5  \\
\hline
\end{tabular}
\caption{\label{tab:param}Propagation parameters consistent with B/C data~\citep{2001ApJ...555..585M}.
The set labelled {\em best} corresponds to the best fit to B/C data, while those
labelled {\em min} and {\em max}  correspond to sets which give minimum and maximum
exotic fluxes~\citep{2004PhRvD..69f3501D}.}
\end{table}
We do not pursue along this line. Rather,
taking advantage of the 1D$\leftrightarrow$2D equivalence, we directly
read these propagation parameters
from the 2D--model, as given in \citet{2001ApJ...555..585M}.
For reasons exposed in Sec.~\ref{sec:1Dlinear-wind},
configurations of maximum wind strength $V_c$ and minimum halo size $L$ lead
to the minimal exotic flux and vice versa.
Table~\ref{tab:param} gathers the set of parameters leading to the maximum/minimum
exotic fluxes, as well as the configuration corresponding to the best fit to B/C.

Using these $max$ and $min$ configurations in Eq.~(\ref{prim:Vc-Diff}) for
the low energy primary exotic flux leads to the result
$N(max)/N(min)=90$. Compared to $\approx 10$ which was obtained from 
the pure-diffusive regime, this implies that $V_c$ increases
the uncertainty on primary fluxes when $\delta$ is not well constrained\footnote{We underline that this
is in fair agreement with the results obtained
i) in a semi-analytical similar---albeit more complicated---study within a 2D propagation
model or ii) using the full propagation scheme in this same 2D
model (see respectively Sec.~III.C and Sec.~IV in \citep{2004PhRvD..69f3501D}).}.
\begin{figure}[!t]
\begin{center}
\includegraphics[width=\columnwidth,angle=0]{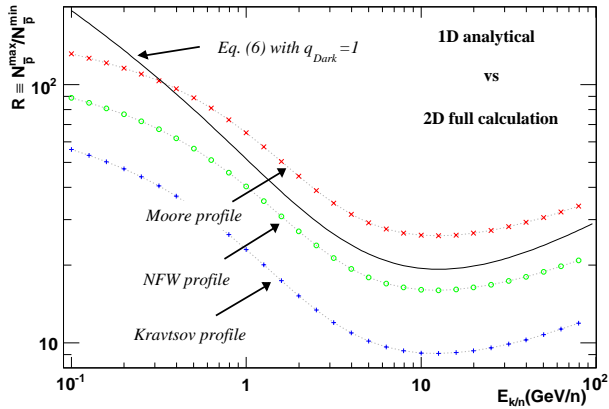}
\caption{\label{fig:error} Ratio $R$ of
the primary fluxes
calculated from the {\em max} and the {\em min} propagation parameter
sets. The solid line results from the 1D modelling Eq.~(\ref{prim:Vc-Diff}),
while all other curves come from full 2D calculations using 
three different dark matter profiles (adapted from~\citep{2005PhRvD..72f3507B}).
}
\end{center}
\end{figure}

This ratio can be estimated at all energies. The corresponding
``energy-dependent" uncertainties can be directly read from
Fig.~\ref{fig:error}, reaching a maximum at low energy and decreasing
at high energy. For comparison purpose, uncertainties from the 2D
modelling are also displayed (using the same propagation parameters;
see details about the 2D--model in the companion paper).
For standard dark matter profiles, the two approaches
are in fair agreement\footnote{The minimum in the curves appear
because propagation models were demanded to fit the 10~GeV/n B/C data point.}.
The various numbers found strengthen the use of the 1D--model for
qualitative estimates of uncertainties.

                     %%%%
               \subsection{How to decrease the astrophysical uncertainties?}
                       \label{subsub:whatif}
The above result is a further support to our approach. It is thus tempting
to pursue along this line and estimate how the uncertainties could
be decreased with better data.

\subsubsection{Sensitivity to the diffusion slope $\delta$}
The uncertainty on the value of $\delta$ greatly enlarges uncertainties
on exotic fluxes at low energy.
\citet{2005APh....24..146C} showed that the ongoing {\sc cream}
experiment~\citep{2004AdSpR..33.1777S} should pinpoint $\delta$
with a precision of $10-15\%$. Reading the corresponding intervals
span by the transport parameters from Figs.~7 and 8 of
\citet{2001ApJ...555..585M}, a simple calculation---analog
to what
is done in the previous subsection---shows that the uncertainty at
low energy is decreased from the previous value of $\sim 100$ down to~$\sim 10$.
The latter value is in agreement with the estimate for fixed~$\delta$, meaning that the
only remaining uncertainty comes now from the $K_0/L$ degeneracy.
The {\sc pamela} experiment~\citep{2006astro.ph..8697P}, which was successfully
launched in June 2006, or {\sc ams} on ISSA 
could as well help determining with a very good precision
this parameter $\delta$.

                       \subsubsection{Radioactive nuclei}
It is well known that radioactive nuclei, e.g. the $^{10}$Be/$^9$Be ratio
lifts the $K/L$ degeneracy.
To complete the exercise, let us investigate to which precision $L$ could be
obtained.
To date, the best data come from {\sc ace}~\citep{2001ApJ...563..768Y}: 
they correspond to a  few energy points only---as it is
the case for almost all radioactive data.
We use the $^{10}$Be$/^9$Be ratio, as heavier isotopes
(such as the $^{26}$Al$/^{27}$Al ratio) suffer larger
experimental uncertainties.

Some quick procedure was exposed in \citet{1979ApJ...229..747J} to extract the
value of $K_0$ (or $L$) and its uncertainties from the data. In our case, as we are only
interested in the ratio of extreme values, i.e. $L^{\rm max}/L^{\rm min}$,
the discussion is even simpler. What is demanded is that
\[
\left(\frac{^{10}{\rm Be}}{^9{\rm Be}}\right)_{\text{\sc ace}} = \left(\frac{^{10}{\rm Be}}{^9{\rm Be}}\right)_{\rm 1D-model}\;.
\]
Expressions for the radioactive species $^{10}{\rm Be}$ and the stable isotope
$^{9}{\rm Be}$ remain to be determined in the 1D framework.
For $^{10}{\rm Be}$, as the decay time $\gamma\tau_0$ is generally much smaller
than the convective, spallative or escape time, the equation to solve in
the 1D--model is simply
\[
-K\frac{d^2N}{dz^2} + \gamma\tau_0 N = 2hq_{\rm sec} \delta(z)
\]
whose solution is
\begin{equation}
\label{eq:rad}
N^{\rm rad}(0)=\frac{hq}{\sqrt{K\gamma\tau_0}}\tanh\left(L
\sqrt{\frac{\gamma\tau_0}{K}}\right)
\approx  \frac{hq}{\sqrt{K\gamma\tau_0}}\;,
\end{equation}
where the approximation comes from $L\sqrt{\gamma\tau_0/K}\gtrsim 10$,
so that $\tanh(\dots)\approx 1$.

For the associated stable isotope, the solution 
can be found in the Appendices. However, it is not even needed: $^{9}{\rm Be}$ is
a stable secondary nuclei having the same propagation history as boron. Demanding
that a model fits B/C---which is a pre-requisite at this stage---automatically ensures the constancy
of the $^{9}{\rm Be}$ flux. It means that
\[
\left\{\frac{\left(^{10}{\rm Be}/^9{\rm Be}\right)^{\rm max}}{\left(^{10}{\rm Be}/^9{\rm Be}\right)^{\rm min}}\right\}_{\text{\sc ace}}
= \left\{\frac{^{10}{\rm Be}^{\rm max}}{^{10}{\rm Be}^{\rm min}}\right\}_{\rm 1D}\;.
\]

Expressing $K=K_0 \beta R^\delta$ and getting the constant ratio $K_0/L$ (fitting B/C) to appear in Eq.~(\ref{eq:rad}),
we obtain
\[
\left\{\frac{^{10}{\rm Be}^{\rm max}}{^{10}{\rm Be}^{\rm min}}\right\}_{\rm 1D}  \approx \sqrt{\frac{L^{\rm max}}{L^{\rm min}}}\;.
\]

Using now for the $^{10}$Be data the $3-\sigma$ error bars given in~\citet{2001ApJ...563..768Y} results in
\[
\frac{L^{\rm max}}{L^{\rm min}}=\left( \frac{0.120+3\times 0.008}{0.120-3\times 0.008}\right)^2=2.25\;.
\]
Hence an uncertainty of about a factor two in $L$ (or equivalently in $K_0$), and the same
uncertainty in the exotic spectra.

However, it should be kept in mind that such a small uncertainty is correct
as long as $\delta$ is known. For instance, \citet{2002A&A...381..539D},
using their derived range of allowed values for $\delta$, found
that the use of radioactive data could not lift the $K_0/L$ degeneracy
(although it slightly reduces it).
Moreover, there could be other issues with radioactive data, as for example
the effect of specific {\em local} features in the solar neighbourhood
(see, e.g., \citealt{2002A&A...381..539D}), which could affect the
determination of the transport parameters.

This exercise has been conducted using existing {\sc ace}
data. Future instruments, such as {\sc ams} have the capability to achieve
better performances.

               \subsubsection{Summary for the constant wind model}
  \begin{table}[!t]
  \begin{tabular}{ c c r c}\hline\\[-8pt]
 \multicolumn{2}{c}{\bf Mean grammage} & \multicolumn{2}{c}{\bf Exotic \boldmath $\bar{p}/\bar{d}$} \\
 \multicolumn{2}{c}{\boldmath$\;\;\;\;\;\langle x\rangle/(\mu v)\;\;\;\;\;$} & \multicolumn{2}{c}{\boldmath$N_{(\bar{p},\;\bar{d})}(z=0)/q_{\rm Dark}$} \\\\[-8pt]\hline\hline
 \multicolumn{4}{c}{ } \\
 \multicolumn{2}{c}{$\displaystyle \frac{1}{2V_c} \cdot \left[1-e^{\left(-\frac{V_c L}{K}\right)}\right]\;$}
  &    \multicolumn{2}{r}{$\;\;\;\displaystyle \frac{K}{V_c^2}\cdot\left[ 1-e^{-\frac{V_cL}{K}} \left(1\!+\!\frac{V_cL}{K}\right)\right]$} \\
 \multicolumn{2}{c}{~~~~\rotatebox{-50.}{$\downarrow$}~~~~~\rotatebox{50.}{$\downarrow$} } & \multicolumn{2}{c}{~~\rotatebox{310.}{$\downarrow$}~~~~~\rotatebox{-310.}{$\downarrow$} } \\
 $\displaystyle \frac{1}{2V_c}$  &  $\displaystyle \frac{L}{2K}$ & $\displaystyle \frac{K}{V_c^2}$ & $\displaystyle \frac{L^2}{2K}$ \\[6pt]
 {$\scriptstyle V_cL/K \gg 1$} & {$\scriptstyle V_cL/K \ll 1$} & {$\scriptstyle V_cL/K \gg 1$} & {$\scriptstyle V_cL/K \ll 1$}\\[-6pt]
 {\tiny (Low E)} & {\tiny (High E)}       & {\tiny (Low E)} & {\tiny (High E)}  \\[6pt]\hline
 \multicolumn{2}{c}{\em Degeneracy}       & \multicolumn{2}{c}{\em Relative uncertainty} \\[2pt]\hline\\[-10pt]
 \multicolumn{4}{c}{\scriptsize \bf  \boldmath{[perfect B/C data: $\delta$} fixed] } \\
    $\times$ & $\mu\cdot L/K_0$ & \hspace{6mm} $\Delta K_0/K_0$ &  $\Delta L/L$ \\[8pt]
 \multicolumn{4}{c}{\scriptsize \bf  \boldmath{[current B/C data$^\dagger$: $\delta\in [0.46-0.85]]$}} \\
 {\footnotesize (from 2D simu.)} &  $\mu\cdot L/K_0$ &  $\sim 100$ &  $\sim 10$ \\[9pt]\cline{2-3}\\[-8pt]
 \multicolumn{4}{c}{\scriptsize \bf  \boldmath{[forthcoming B/C data + old $^{10}$Be/$^9$Be}]} \\
 \multicolumn{2}{c}{\footnotesize $\rightarrow$ degeneracy broken} &  $\sim 2^\ddagger$ &  $\sim 2^\ddagger$ \\[5pt]\hline\hline
       \end{tabular}
{~~~\scriptsize $^\dagger$ Unchanged whether radioactive nuclei are considered or not.\\}
{\scriptsize $^\ddagger$ Astrophysical uncertainty should vanish using new radioactive data.}
\caption{\small Summary of results derived in Secs.~\ref{sec:diff} and~\ref{sec:1Dlinear-wind}.
The first row corresponds to (1) left column: formula for the grammage fitted from B/C data and
(2) right column:
the primary flux when spallative destruction is discarded. After
the second row, each column (1 and 2)
is further subdivided in two parts corresponding to the asymptotic behavior
i) at low energy (left; convection-dominated) and ii) at high energy (right; diffusion-dominated).
Note that the convection-dominated regime is not necessarily reached.
In the subsequent rows, col.~(1) corresponds to the degeneracies (derived from
the above formulae) in the transport parameters and col.~(2) to the uncertainties
associated with these degeneracies. The same structure (further separation into low and high energy) is kept
as for the upper half of the table. For example, for current B/C data at high energy, the degeneracy
is $\mu\cdot L/K_0$, leading to a relative uncertainty of $\Delta L/L$
for exotic species.
 \label{tab:vc}}
       \end{table}

Table~\ref{tab:vc} summarizes the qualitative results derived in this section.
Determining the slope of the diffusion coefficient
from B/C data should be the primary concern, whereas radioactive nuclei
ultimately lift the remaining degeneracies.

We also remind that, although it was not explicitly stated all along,
all the reasonings do hold for exotic anti-protons as well as
for exotic anti-deuterons.

%/////////////////////////////////////////////////////////////////
\section{Summary and conclusions}
\label{sec:theEnd}

We used 1D--propagation models to study astrophysical uncertainties on
exotic \pbar\ and \dbar\ fluxes (e.g. from annihilation of dark matter
in the Galactic halo). 
We started with a detailed analysis of the constant wind
diffusion model to recall and understand the origin and key parameters
of the uncertainties for exotic fluxes.

We then stressed the origin of uncertainties for the transport parameters in
the constant and linear wind models, and their differences at low energy,
emphasizing on how crucial the determination of the diffusion slope $\delta$ was.
We showed that the present astrophysical uncertainties of ${\cal O} (100)$
could be severely reduced with a good determination of $\delta$.

We finally showed that even if the internal degeneracies of a given model
are reduced, the differences between different modellings (e.g. assuming
different spatial dependences of some of the transport coefficients)
could be important. For the constant and the linear wind model, a lower
limit of $30\%$ was derived. These differences between models are bound
to become a more serious matter in the near future.

%****************************************************************************
%****************************************************************************
%****************************************************************************

%****************************************************************************
%****************************************************************************
%****************************************************************************
\section*{Acknowledgments}
D.~M. warmly thanks P. Salati and all the participants
of the Predoctoral summer school of Les Houches 2005 for the
very pleasant and exhilarating fortnight. Some of the material
presented ihere is based on notes which were written for
a lecture at this school.

%****************************************************************************
%****************************************************************************
\appendix
\section{Reminder of the equivalence between Leaky Box and diffusion models}
\label{app:LBvsDM}
The link between diffusion models and Leaky Box Models is established through
the weighted-slab formalism. The weighted-slab
decomposition is only approximate when energy losses
are taken into account \citep{1979Ap&SS..63..279L,1991ICRC....2..268J}.
But, in this paper, this is not an issue as they are
neglected for all 1D models.

                         %%%%
               \subsection{The weighted-slab formalism in brief}
The ansatz is to rewrite
the differential density $N(\vec{r})$ of any transport equation
(e.g. Eq.~\ref{1D-generic}) as $N(\vec{r})=\int_{0}^{\infty}\tilde{N}(x)G(x,\vec{r})dx$
(where $x$ is the grammage crossed by a CR in g~cm$^{-2}$).
Demanding $\tilde{N}(x)$ to follow the slab equation, one is
left with an equation on $G(x,\vec{r})$. The slab equation
is common to all models (LB, 1D and 2D--diffusion models, etc.)
and describes the nuclear aspects of propagation.
The remaining equation only depends on the transport properties and on
the choice of the geometry, both assumed to be independent of the species
(see Sec.~2 of \citealt{2001ApJ...555..585M}, for a longer albeit succinct
summary; or \citealt{1990acr..book.....B}, p.~45 for a more complete treatment).
The exercise is easily done on the leaky box equation (no spatial dependence)
governing the evolution of the nucleus $j$ (source term $q^j$ plus
secondary contributions from all heavier nuclei $k$)
       \begin{equation}
       \bar{N}/\lambda_{\rm esc}(E) + \frac{\sigma^j}{\bar{m}} \bar{N^j}=
       \bar{q^j}+\!\!\sum_{k>j}\!\frac{\sigma^{kj}}{\bar{m}}\bar{N^k}\;\;.
       \end{equation}
Using the ansatz and inserting the slab equations
       \begin{equation}
       \frac{d N^j(x)}{d x}+ \frac{\sigma^j}{\bar m}
                       N^j(x) = \bar{q^j} +
                       \sum_{k=j+i}^{n_{max}}\frac{\sigma^{kj}}{\bar m}N^k(x)
       \end{equation}
leads to an equation for $G(x)$ whose solution is given by
       \begin{equation}
       \label{PLD-LB}
    G^{LBM}(x)=\frac{1}{\lambda_{\rm esc}}\exp \left(
                \frac{-x}{\lambda_{\rm esc}}\right)\;.
       \end{equation}
Different propagation models can be compared through their path length
distributions at the location where CR measurements are made, i.e.
$\vec{r}_\odot$. Thus, a diffusion models (DM) will be locally equivalent
to a LBM if
       \begin{equation}
       \label{PLD-equiv}
   G^{DM}(x,\vec{r}_\odot)=G^{LBM}(x)\;.
       \end{equation}
We remind that for a LBM all locations come to the same thing since
all quantities are spatially averaged.

                         %%%%
               \subsection{PLD and mean grammage $\langle x \rangle$}
               \label{app:LBvsDM-2}
A useful quantity is the mean grammage crossed by CRs, defined as
       \begin{equation}
       \label{<x>1}
               \langle x\rangle (\vec{r}) = \frac{\int_0^{\infty} xG(x,\vec{r})dx}{\int_0^{\infty} G(x,\vec{r})dx}\;.
       \end{equation}
An alternative expression is provided by
       \begin{equation}
       \label{<x>2}
    \langle x\rangle(\vec{r})= -\bar{m}\left.\left( \frac{d}{d\sigma^p}\ln
                N^p(\vec{r}) \right)
                \right|_{\sigma^p=0}\;\;,
       \end{equation}
where $N^p(\vec{r})$ is the solution for a primary species
(see \citealt{1990acr..book.....B}, p.~53).
The advantage of this second formulation is that there is no need
to know the PLD, $G(x,\vec{r})$, to evaluate the mean grammage.
Indeed, the former quantity can be trickier than $N^p(\vec{r})$ to derive
(as an illustration, PLDs for several modellings can be found, e.g.,
in \citet{1976Ap&SS..40..357O}).

The mean grammage is only the first moment of the distribution
$G(x,\vec{r})$: only the latter contains the full
information on the transport for the studied model.
However, for two propagation models to be equivalent, we
will use a weaker version of Eq.~(\ref{PLD-equiv}), only
demanding the equivalence of the first moments
       \begin{equation}
       \label{x-equiv}
    \langle x\rangle^{DM}(\vec{r}_\odot)=\langle x\rangle^{LBM} \equiv
                \lambda_{\rm esc}(E)\;.
       \end{equation}
The last equality comes from Eq.~(\ref{PLD-LB}) plugged into Eq.~(\ref{<x>1}).
We remind that all the above quantities depend on the energy. Note also
that this relation does imply equivalence of PLDs for many
cases especially if the thin-disk approximation is assumed
(see App.~\ref{App:mean_grammage}).

\section{Mean grammage $\langle x \rangle$ in two modellings}
\label{App:1D-standard}
The diffusion coefficient $K$ is the same in the disk and in the halo
(see Fig.~\ref{fig:Lbound2})\footnote{Contrarily to an erroneous calculation
in the previous versions of this paper, in the thin disk limit,
setting two distinct diffusion coefficients for the disk and for the
halo is strictly equivalent to have only one diffusion coefficient for the two zones.
Indeed, in the thin disk approximation, the time spent in the disk tends to zero,
while the number of crossing (hence the grammage) is kept constant: only the diffusion
coefficient in the halo plays a role. This result is of course known for
almost as long as diffusion models exist. The trivial mistake we made (shame on us...
and oh, yes, a referee to point out the mistake would have been just the right thing
to have, but as refering goes, you know...) was connected to a wrong implementation of the continuity
condition at the crossing of the zones. A very small part of the paper was affected by this mistake,
and the only one plain wrong conclusion is now removed.}.
The galactic wind is taken either as a constant $V_{\rm gal}(z)=V_c$ or as
a linear term $V_{\rm gal}(z)=V_l \times z$.
The transport equation reads:
\begin{equation}
\label{1D-general-thin-disk}
 -KN''+(V_{\rm gal}N)'+2h\delta(z).nv\sigma \times N = 2hq\delta(z).
\end{equation}
The quantity $n\sigma v$ is the total destruction rate of the
anti-nucleus under study with the interstellar gas, whose density is $n$.

 \subsection{Solutions}
For our purpose, it is sufficient to extract solutions for
primary standard species, that lead straightforwardly
to the mean grammage $\langle x \rangle$.

                           %-------%
               \subsubsection{Constant wind $V_{\rm gal}(z)=V_c$}
               \label{App:general-Vc}
We start with the equation in the galactic halo, i.e.
\[
     - KN'' + V_c N' =0 \quad \rightarrow  \quad N=Ae^{V_cz/K}+B\;.
\]
Implementing the boundary condition $N(z=L)=0$ determines one
of the two constants of the solution in the halo:
\[
      N=A \left( e^{V_cz/K}- e^{V_cL/K}\right) \;.
\]

We integrate over the thin-disk (see e.g.~\citep{2001ApJ...547..264J}),
i.e. $\lim_{\epsilon\rightarrow 0}\int_{-\epsilon}^{+\epsilon}$(Eq.~\ref{1D-general-thin-disk})$dz$,
taking care of the discontinuity induced by the wind during the disk crossing:
\[
 -2K N'(0)+ 2V_cN(0) +2hnv\sigma N(0) = 2hq\;.
\]
The solution for the halo $N$ above is plugged into this equation.
It is convenient for the following to write the solution as:
\begin{equation}
N(z)=N(0)\times \frac{1-e^{-\frac{V_c}{K}(L-z)}}{1-e^{-\frac{V_c}{K}L}}
\end{equation}
with
\begin{equation}
\label{eq:final-Vc}
N(0)=\frac{2hq}{2V_c/(1-e^{-V_cL/K})+ 2hnv\sigma}\;.
\end{equation}

                           %-------%
               \subsubsection{Linear wind $V_{\rm gal}(z)=V_l\times z$}
               \label{App:general-Vl}

The parameter $V_l$ has the dimension of the inverse of a time.
As above, we start by solving this equation in the halo, where
it reduces to
\begin{displaymath}
     - KN'' + V_l z N' + V_l N =0\;.
\end{displaymath}
This equation is not hermitian
and it is more convenient to use new variables $n$ and $y$
defined as
\begin{displaymath}
 n(y)=N(y) \, e^{y^2}
 \quad \text{and} \quad
 y \equiv \sqrt{2}\nu z
\end{displaymath}
where
\begin{equation}
\label{eq:nu}
       \nu\equiv \sqrt{\frac{V_l}{2K}}\;.
\end{equation}
The diffusion equation then reads
\begin{displaymath}
 n''(y) - n(y)  \left(\frac{1}{2} + \frac{y^2}{4} \right) = 0\;.
\end{displaymath}
The general solution of this equation is given in
\citet{1994toi..book.....G} (p. 1067, Eq.~9.255). Rewritten
using $z$ gives
\begin{displaymath}
N(z) = A e^{(\nu z)^2} + B e^{(\nu z)^2} {\rm erf}(\nu z).
\end{displaymath}
The constants of integration $A$ and $B$ are determined as usual.
First, the condition $N(z=L)=0$ yields\footnote{The function $er\!\!f$ is defined
as $\displaystyle {\rm erf}(z)\equiv\frac{2}{\sqrt{\pi}}\int_0^ze^{-t^2}dt$.}
\begin{displaymath}
 A = - B \times {\rm erf}(\nu L).
\end{displaymath}
Second, integration of Eq.~(\ref{1D-general-thin-disk})
over the thin disk yields
\begin{displaymath}
 -2 K N'(0) + 2h \Gamma  N(0) =  2h q \;,
\end{displaymath}
which determines $A$ and gives the final expression
\begin{equation}
 N(z) = N(0)\times  e^{(\nu z)^2} \;
       \left(
       \frac{{\rm erf}(\nu L) - {\rm erf}(\nu z)}{{\rm erf}(\nu L)}
       \right)
\end{equation}
with
\begin{equation}
\label{eq:final-Vl}
 N(0) = \frac{2hq}{\frac{4\nu K}{\sqrt{\pi}{\rm erf}(\nu L)} + 2hnv\sigma}
               \;,
\end{equation}
where $\nu$ is defined in Eq.~(\ref{eq:nu}).

                  %%%%%%%%%%%%%%%%%%%
               \subsection{Mean-grammage $\langle x\rangle$ at $z=0$}
               \label{App:mean_grammage}
Two approaches can be used to derive the mean grammage.
The first one is depicted in App.~\ref{app:LBvsDM}.
The second one is based on the
fact that, in the thin disk approximation, all solutions
for $z=0$ can be recast to look formally as to a Leaky Box solution
(see \citealt{2001ApJ...547..264J}):
\begin{equation}
       N(0)^{\rm 1D-model} = \frac{2hq/\mu v}{1/\langle x\rangle^{\rm 1D-model} +
       \sigma/\bar{m}}\;.
\end{equation}

In that case, demanding an equivalence between the mean grammages
of different models (as in App.\ref{app:LBvsDM-2})
is one and the same to demanding the equivalence between
path length distributions. The corresponding mean-grammages are
directly obtained from Eqs.~(\ref{eq:final-Vc}) and (\ref{eq:final-Vl}).
Using the relation $\mu\equiv 2hn\bar{m}$ ($\bar{m}$ is the mean mass
of the interstellar medium and $\mu$ is the surface mass density
of the gas in the disk),
\begin{itemize}
 \item Constant galactic wind $V_{\rm gal}(z)=V_c$
               \begin{equation}
                \label{x-1D-Vc-thin}
                               \langle x\rangle^{V_c}= \frac{\mu v}{2V_c}\times
                       \left(1-e^{(-V_cL/K)}\right)\;.
               \end{equation}
 \item Linear galactic wind $V_{\rm gal}(z)=V_l \times z$
               \begin{equation}
                \label{x-1D-Vl-thin}
                       \langle x\rangle^{V_l}= \frac{\mu v \times \sqrt{\pi} {\rm erf}
                        \left( L\sqrt{V_l/(2K)}\right)}{4K\sqrt{V_l/(2K)}}\;.
               \end{equation}
\end{itemize}
Note that both formulae reach the limit of pure diffusive transport
for vanishing (compared to the diffusion
coefficient) values of the wind :
\begin{equation}
       \label{x-1D-pure-thin}
               \langle x\rangle^{\rm pure-diffusion} = \frac{L\mu v}{2K}\;.
\end{equation}

%****************************************************************************
\section{\pbar\ and \dbar\ fluxes from sources in the diffusive halo}
\label{App:1D-exotic}
We now use a constant source term in the Galaxy:
\begin{equation}
\label{1D-exotic-thin-disk}
 -KN''+(V_{\rm gal}N)' +n\delta(z).v.\sigma \times N = q.
\end{equation}

               \subsection{Constant wind $V_{\rm gal}(z)=V_c$}
We proceed as for the standard source case. In the halo,
the transport equation reads
\[
 -KN''+ V_c N' = q\;.
\]
Using the boundary condition $N(z=L)=0$, gives
\[
N(z)=A\times \left(\frac{1-e^{-\frac{V_c}{K}(L-z)}}{1-e^{-\frac{V_c}{K}L}}
\right) +\frac{q}{V_c}(z-L)\;.
\]
Integration on the thin-disk gives the relation
\[
-2K N'(0) + 2V_cN(0)+2hnv\sigma N(0) =0\;,
\]
so that the final solution is
\begin{eqnarray}
   \left\{
   \begin{array}{ll}
\label{eq:prim-Vc-general}
       N(z)=\frac{qL}{V_c}
\left\{
 \frac{(1+\alpha+\xi) \cdot(1-e^{\frac{-V_c}{K}(L-z)})}{
 (\alpha+\xi+\frac{\alpha e^{-\alpha}}{1-e^{-\alpha}}) (1-e^{-\alpha})}
 +\frac{z}{L}-1\right\}\vspace{4mm}
 \\
\displaystyle N(0)=\frac{qL}{V_c}\left\{
\frac{1-(1+\alpha)e^{-\alpha}}{(\alpha+\xi)(1-e^{-\alpha}) +\alpha e^{-\alpha}}
\right\}
   \end{array}
   \right.
\end{eqnarray}
where
\begin{equation}
       \alpha \equiv \frac{V_c L}{K} \quad {\rm and} \quad
       \xi \equiv \frac{hnv\sigma L}{K}.
\end{equation}

               \subsection{Linear wind $V_{\rm gal}(z)=V_l\times z$}
                       \label{sec:prim-Vl}
In the halo, the equation is
\begin{displaymath}
     - KN'' + V_l z N' + V_l N =q\;.
\end{displaymath}
Using $\nu\equiv \sqrt{V_l/(2K)}$ and following closely
the derivation of Sec.~\ref{App:general-Vl},
we find the solution
\begin{displaymath}
N(z) = A e^{(\nu z)^2} + B e^{(\nu z)^2} {\rm erf}(\nu z)+\frac{q}{V_l}.
\end{displaymath}
The condition $N(z=L)=0$ yields
\begin{displaymath}
 A = - B \times {\rm erf}(\nu L) -\frac{q}{V_l}e^{-(\nu L)^2}.
\end{displaymath}
Integration of Eq.~(\ref{1D-exotic-thin-disk}) over the thin-disk yields
\begin{displaymath}
 -2 K N'(0) + 2h nv\sigma  N(0) =0\;,
\end{displaymath}
which determines $A$ and gives the final expression
\begin{eqnarray}
   \left\{
   \begin{array}{ll}
\label{eq:prim-Vl-general}
\displaystyle
N(z)\!=\!\frac{q}{V_l}
\left\{
 1-\frac{e^{-(\nu L)^2}}{e^{-(\nu z)^2}}
       \left[
        1 - \frac{{\rm erf}(\nu z)-{\rm erf}(\nu L)}{\frac{2\nu L}{\xi\sqrt{\pi}}+{\rm erf}(\nu L)}
       \right]
\right\}
\vspace{4mm}
\\
\displaystyle
N(0)\!=\!\frac{q}{V_l}
\left\{
  1\!-\!e^{-(\nu L)^2}
       \left[
         1\!+\!\frac{\xi\sqrt{\pi}\;{\rm erf}(\nu L)}{2\nu L + \xi\sqrt{\pi}\;{\rm erf}(\nu L)}
       \right]
\right\}
   \end{array}
   \right.
\end{eqnarray}
where
\begin{equation}
       \nu\equiv \sqrt{\frac{V_l}{2K}}
\quad {\rm and} \quad
       \xi \equiv \frac{hnv\sigma L}{K}.
\end{equation}

               \subsection{Formulae for $\xi=0$ (no spallations)}
\begin{itemize}
       \item Constant wind $V_c$
               \begin{eqnarray}
           \label{prim:pureDiff-nospal_z}
   \left\{
   \begin{array}{ll}
       N(z)=\frac{qL}{V_c}
\left\{ (1+\frac{K}{V_cL})\left(1-e^{-\frac{V_c}{K}(L-z)}\right)+\frac{z}{L}-1
\right\}\;,\vspace{2mm}
 \\ \displaystyle
       N(0)=\frac{qK}{V_c^2}
\left\{ 1-\left(1+\frac{V_cL}{K}\right)e^{-\frac{V_cL}{K}}\right\}\;.
   \end{array}
   \right.
               \end{eqnarray}
       \item Linear wind $V_l$
               \begin{eqnarray}
           \label{prim:Vl-nospal}
           \left\{
           \begin{array}{ll}
               \displaystyle N(z)= \frac{q}{V_l}
                                 \left(1-e^{-\frac{V_l}{2K}(L^2-z^2)}\right)\;,\\
               \displaystyle  N(0)= \frac{q}{V_l}
                                 \left(1-e^{-\frac{V_lL^2}{2K}}\right)\;\;.
           \end{array}
           \right.
               \end{eqnarray}
\end{itemize}

Finally, at sufficiently high energy, the purely diffusive
regime is reached, leading, for both wind configurations, to
       \begin{eqnarray}
   \label{prim:pureDiff-nospal}
   \left\{
   \begin{array}{ll}
       \displaystyle N(z)= \frac{q(L^2-z^2)}{2K}\;,\\
       \displaystyle  N(0)= \frac{qL^2}{2K}\;\;.
   \end{array}
   \right.
       \end{eqnarray}

       \section{The impact of neglecting spallations}
\label{app:nospal}
Solutions switching-on/off spallations for the constant wind model
are compared using the master Eq.~(\ref{eq:prim-Vc-general}).
\begin{figure}[!t]
\begin{center}
\includegraphics[width=.65\columnwidth,angle=0]{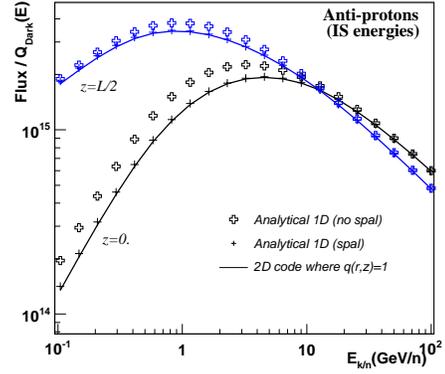}
\caption{\label{fig:spal_on_off}Exotic \pbar\ flux for $K_0=1.35\times 10^{28}$~cm$^2$~s$^{-1}$ (and $\delta=0.7$),
$L=4$~kpc and $V_c=12$~km~s$^{-1}$. Symbols correspond
to the 1D model with (plus) or without (open plus) spallative destruction. This is shown for
an observer in the disk ($z=0$) or at $z=2$ kpc.}
\end{center}
\end{figure}
Figure~\ref{fig:spal_on_off} displays a comparison of \pbar\ fluxes 
for the best fit propagation parameters.
The inclusion of spallations for \pbar\ is of little consequence. We
checked that it remains the case for other propagation parameters
compatible with B/C data. The most important effect is when $V_c=0$ (which
is usually never met). In that case, the maximum deviation is observed at low energy and
is about several tens of percent---but we remind that due to the Solar modulation, fluxes below
$\sim 500$~MeV/n IS energies are not reachable. The deviation is expected to be
slightly larger for
$\bar{d}$, whose destruction cross section is roughly twice the \pbar\ one.
For consistency, the 1D results were also compared with 2D calculations (see companion) using $q_{Dark}(r,z)=1$.
\vspace{-1.5cm}
%%%%%%%%%%%%%%%%%%%%%%%%%%%%%%%%%%%%%%%%%%%%%%%%%%%%%%%%%%%%%%%%%%%%%%%%
%%%%%%%%%%%%%%%%%%%%%%%%%%%%%%%%%%%%%%%%%%%%%%%%%%%%%%%%%%%%%%%%%%%%%%%%
\bibliography{mtc.bib}
\end{document}